\documentstyle[preprint,aps,epsf]{revtex}
\begin{document}

\title{
QUANTUM DOT LATTICE EMBEDDED IN AN ORGANIC MEDIUM:HYBRID EXCITON STATE
AND OPTICAL RESPONSE}

\author{
{\bf 
Nguyen Que Huong and Joseph L. Birman} \\
Physics Department, The City College, CUNY\\
Convent Ave. at 138 St, New York, NY 10031, USA}

\date{\today}
\maketitle

\begin{abstract}

We propose a new model to implement organic exciton - semiconductor 
exciton hybridization by embedding a semiconductor quantum dot array into 
an organic medium. A Wannier Mott transfer exciton is formed when the 
exciton in each semiconductor dot interacts via multipole-multipole 
coupling with other excitons in the different dots of the array. A new 
hybrid exciton appears in the system owing to strong dipole-dipole 
interaction of the Frenkel exciton of the organic molecules with the 
Wannier Mott transfer exciton of the quantum dot array. This hybrid 
exciton has both a large oscillator strength (Frenkel like) and a large 
exciton Bohr radius (Wannier like). At resonance between these two types 
of excitons, the optical non-linearity is very high and can be controlled 
by changing parameters of the system such as dot radius and dot spacing.

\bigskip
\noindent

\bigskip
PACS numbers: 73.20.Dx, 78.60.J, 42.65, 71.35  
\end{abstract}
\newpage
{\bf I.Introduction.}

The possibilities of using organic materials to embed semiconductor 
heterostructures lead us to the expectation of having some system with new 
unique physical properties. 
It has already been noted that a new type of excitonic state could be 
obtained by mixing Wannier-Mott and Frenkel excitons.
In [1-4] the authors proposed structures of layers with 
organic-inorganic quantum 
wells, and parallel organic-inorganic quantum wires, where the Wannier and 
Frenkel 
excitons in neighboring layers interact with each other to form a new 
type of exciton, which has both the very large oscillator strength of the 
Frenkel exciton, and the high polarizability in an external field of a  
Wannier exciton. Recently, in [3] the interesting model of a 
single semiconductor quantum dot with an organic shell was also 
proposed and 
the strong enhancement of the non-linearity was found for the weak 
confinement 
regime in the limit of dot radius $R_D>>a_B$ (exciton Bohr-radius).

At present, quantum dots - which are the three-dimensionally  confined 
nanostructures - have been studied intensively theoretically and 
experimentally [5-9]. As a new model for the 
heterostructures, where the resonant mixing of 
Wannier-Mott and Frenkel exciton can appear, we propose here the system of 
a semiconductor quantum dot array embedded in a medium of organic 
material. 
Such structures were reported to have been fabricated in several labs 
[10, 16]. 
It is already known [11] that when many quantum dots are arranged 
together in an array, due to the multipole interaction of excitons in 
different dots, an  exciton inside a quantum dot can be considered 
as not localized in that dot, but it can propagate through the lattice via 
the mechanism of exciton transfer processes. 
In our model, when we place such a dot array in an organic material, due to 
the 
interaction of this propagating exciton with  the  medium, a 
new hybrid exciton will appear in a system. 
This hybrid exciton, which is a mixed
state of the transfer exciton and the Frenkel exciton of the medium when 
these are at  resonance, also has a large exciton radius ( because of the 
large Wannier-Mott  exciton radius) and a large oscillator strength 
(because of the large oscillator strengths of the both Frenkel and 
transfer Wannier-Mott excitons).
The small mass of the transfer exciton then leads to a  large coherence 
length of the system. 
This fact, as well as the hybridization, will give us a very large 
optical non-linearity.

In section II we present our model to get the hybrid exciton Hamiltonian. 
The hopping coefficient of the exciton in different dots and the Wannier- 
Frenkel exciton coupling constant are described in section III and IV. 
The nonlinear optical susceptibility and its dependence upon the system 
parameters are obtained in section V, along with an estimate of the 
magnitude of the non-linear coefficient for a typical system.

{\bf II. The Model.}

We consider a three dimensional array of semiconductor quantum dots 
placed into some organic material as a host medium. 
The size of the system should be considerably smaller than the 
wavelength of light that corresponds to
the transition between excited and ground state [3]. 
As an approximation, we will consider here the ideal array with dots of 
the same radius $R$ and the same dot-dot spacing $d$. 

The total Hamiltonian of the system will be taken as follows: 
\begin{eqnarray}   
H & = & \sum_{{\vec n},l}{E^{W}_{{\vec n}l}a^{+}_{{\vec n}l}a_{{\vec 
n}l}} +  
\sum_{{\vec k},m}{E^{F}_{m}({\vec k})b^{+}_{{\vec k}m}b_{{\vec k}m}} 
\nonumber \\
& & + \sum_{ml{\vec n}{\vec k}}{g_{lm}({\vec r_{\vec n}},{\vec 
k})(a^{+}_{{\vec n}l}b_{{\vec k}m} + a_{{\vec n}l}b^{+}_{{\vec k}m})} + 
\sum_{{\vec n}{\vec n'}ll'}{t_{{\vec n}{\vec n'}ll'}(a^{+}_{{\vec 
n}l}a_{{\vec n'}l'} + h.c.)} 
\end{eqnarray}

\noindent
where $a^{+}_{{\vec n}l}$, ($a_{{\vec n}l}$) are creation (annihilation) 
operators of Wannier excitons in quantum dots. 
Index $l$ labels the exciton states and 
${\vec n}$ indicates the sites of the dot in the dot lattice. 
Here we assume that the dots 
are distributed on sites of a  three-dimensional lattice with the position  
${\vec n}= (n_x, n_y, n_z)$ of each site in (x,y,z) coordinates, where the 
distance between the sites (the "lattice" constant) equals d. 
For simplicity we assume a cubic array, i.e, the number of dots in each 
directions $ N_x, N_y, N_z $ is the same $ N_x = N_y = N_z = N $. 
$b^{+}_{{\vec k}m}$, ($b_{{\vec k}m}$) 
are creation (annihilation ) operators for the Frenkel exciton in the 
organic medium with wave vector ${\vec k}$ in the $m^{th}$-exciton state. 
$E^{W}_{{\vec n}l}$ and 
$E^{F}_{{\vec k}m}$ are the excitation energies of Wannier excitons in the 
dots and the Frenkel exciton in the medium, respectively. For the Wannier 
excitons confined to a dot, or quantum sphere, the
oscillator strengths are concentrated mainly on the low excited states. 
So, with no loss of generality, and in order to simplify we will
consider only the interaction of the lowest states of excitons( the
ground state and the lowest excited state). Also we
assume that the energy difference between the energy levels $E^{F}$ and
$E^{W}$ is much smaller than the distance to other bands.
$g_{lm}({\vec r_{\vec n}},{\vec k})$ is the coupling constant of Wannier and 
Frenkel excitons, and $t_{{\vec n}{\vec n'}ll'}$ is the hopping constant 
between Wannier excitons in the dots. 
This hopping constant, which has its origin in the 
multipolar interaction of excitons in different dots, in general, is 
different in different directions because of the direction of 
polarization. Here we assume only nearest dots interact with each other, 
so the hopping constants for the nearest dots in the x, y, z direction are 
$t_x, t_y, t_z$, respectively, where e.g. $ t_x = t_{{\vec n}{\vec 
n'}ll'}$ if ${\vec n'}$ is $(n_x+1, n_y, n_z)$ and $l, l'$ are the label 
of the lowest energy levels, and similarly for $t_y, t_z$. 
Because the dot radius is approximately equal to the ground state exciton 
Bohr radius for systems under consideration, we assume there exists 
only one exciton in each dot, so we omit the exciton-exciton 
interaction in the same dot.
Then we have for the hopping term:
\begin{eqnarray}
H_{hop} & = & 
\sum_{ll'}{\sum_{n_x,n_y,n_z}{(t_xa^{+}_{n_x+1,n_y,n_z;l}a_{n_x,n_y,n_z;l'} + 
t_ya^{+}_{n_x,n_y+1,n_z;l}a_{n_x,n_y,n_z;l'}}} +\nonumber \\ 
& & + t_za^{+}_{n_x,n_y,n_z+1;l}a_{n_x,n_y,n_z;l'} +h.c. ) 
\end{eqnarray}

\noindent
Changing to k-space by Fourier transformation, we obtain:
\begin{equation}
a^{+}_{n_x,n_y,n_z;l} =1/{\sqrt{N^3}} \sum_{k_x,k_y,k_z}{[\exp{i(n_xk_xd + 
n_yk_yd + n_zk_zd)}]a^{+}_{k_x,k_y,k_z;l}}
\end{equation}

\noindent
Here d is the distance between dots, ${\vec k}= \{k_x,k_y,k_z\}$ is the wave 
vector of the exciton in the coupled dots.

The Fourier transform of $t_x, t_y, t_z$ will be $t(k_x), t(k_y), 
t(k_z)$, respectively.

The Fourier transformation for the dot-medium coupling constant 
$g_{lm}({\vec r_{\vec n}},{\vec k})$ will be the following:
\begin{equation}
g_{lm}({\vec r_{\vec n}}, {\vec k}) =\sum_{\vec {k'}}{[\exp{i(n_xk'_xd + 
n_yk'_yd + n_zk'_zd)}]G_{lm}({\vec k'},{\vec k})}.
\end{equation}

\noindent
Notice here that if one makes the translational transformation with a 
lattice vector ${\vec L}$ in the lattice, due to the exponential forms of 
the Frenkel and Wannier exciton state functions, one will get, by translational 
invariance, $\sum_{L}{\exp{i({\vec k}-{\vec k'}){\vec L}}} =\delta 
(k-k')$. So the 
coefficient $G_{lm}({\vec k},{\vec k'})$ will be different from zero only 
if $ k = k'$. Then instead of $G_{lm}({\vec k},{\vec k'})$ we can write 
the coupling constant as  $G_{lm}({\vec k})$ and omit the sum over $k'$.

Then we get the total Hamiltonian (1) as:
\begin{eqnarray}
H & = & \sum_{l{\vec k}}{E^{W}_l({\vec k})a^{+}_{l{\vec k}}a_{l{\vec k}}} +
2\sum_{l}{\sum_{k_x,k_y,k_z,l}{(t(k_x)\cos{k_xd}
+ t(k_y)\cos{k_yd} + t(k_z)\cos{k_zd})a^{+}_{l{\vec k}}a_{l{\vec k}}}} 
\nonumber \\ & & + \sum_{m{\vec k}}{E^{F}_m({\vec k})b^{+}_{{\vec 
k}m}b_{{\vec k}m}}  + \sum_{ml{\vec k}}{G_{lm}({\vec 
k})(a^{+}_{{\vec k}l}b_{{\vec k}m} + a_{{\vec k}l}b^{+}_{{\vec k}m})}
\end{eqnarray}

\noindent
In the perfect confinement
approximation [1,4], the exciton wave functions must vanish at the 
boundary of 
dots, and the energy of the Wannier exciton in the spherical quantum dot 
has discrete values according to the zeros of the Bessel function. 
That is, the energy of the Wannier exciton with quantum number $\{n,l\}$ is 
$E^{W}_{nl}(k) = E_g - E^b_{ext} + \frac{\hbar^2\gamma^2_{nl}}{2mR^2}$, 
where $E_g$ is the band gap, $E^b_{ext}$ is the exciton
binding energy, and $\gamma_{nl}$ is the n-th zero of the spherical 
Bessel function  $J_l(x)$ of order $l$, which depends on the magnitude of 
the dot radius R. 
Here m is the effective mass of the electron and the hole. 
The lowest excitation in the dot will be the state with $l=0, n=1$. In 
writing the expression for $E^{W}_{nl}(k)$ we are assuming the dot is 
spherical as a good approximation to the actual shape. 

For kd small, the total Hamilton (5) can be written as: 
\begin{equation} 
H  = \sum_{{\vec k}}{E^{W}_{tr}({\vec k})} + \sum_{k}{E^{F}(k)b^{+}_{{\vec 
k}}b_{{\vec k}}} + 
\sum_{k}{G(k)(a^{+}_{\vec k}b_{\vec k} + a_{\vec k}b^{+}_{\vec k})}
\end{equation}

\noindent 
where $E^W_{tr}({\vec k})$ is the transfer energy of the Wannier exciton 
of the semiconductor dot array:  
\begin{equation} 
E^W_{tr}({\vec k}) = E^{W} + 2\sum_{i}{t(k_i)} - d^2\sum_{i}{t(k_i)k^2_i},
\end{equation}
 
\noindent 
where $i = \{x,y,z\}$.
  
Equation (6) describes the system of Wannier excitons and Frenkel excitons, 
interacting with each other. 
Besides the exciton energy in single quantum dots, the energy of the Wannier 
exciton in the quantum  dot array also includes the large transfer energy 
between two nearest quantum dots in the array. 

We notice that because of
confinement, the energy and the state of one quantum dot cannot be described by the wave vector. 
But the energy and the state of the transfer exciton in the quantum dot array does have the k-vector
dependence and we will need to include the dispersion relation for the energy of this transfer exciton. 
This energy strongly depends on the value of the hopping constant $t(\vec 
k)$ and the direction of the
polarization vector of the exciton, which we will investigate in the next 
sections.  The presence of the transfer exciton allows us to change 
the energy region of the resonance and also the optical properties of the 
hybrid exciton. So in our model the
semiconductor quantum dot array embedded in an organic medium can be interpreted as follows: the Wannier
excitons in quantum dots interact with each other to form a kind of 
transfer exciton propagating
through the lattice. This transfer exciton in its turn is coupled at 
resonance with the Frenkel
excitons in the organic medium to form an hybrid organic-inorganic exciton state.

In our model we consider the exciton-exciton interaction and the 
hybridization as the principal 
effect, so here we omit the potential scattering between the dot array and 
the medium and leave it for future reseach. 

We write the new hybrid excited state as the following: 
\begin{equation} 
|\Psi(k)> = u_l(k)f^F(0)\Psi^{W}_l(k) + v_{l'}(k)f^W(0)|\Psi^{F}_{l'}(k)> 
\end{equation}

\noindent 
where $\Psi^{W}(k)$ and $\Psi^{F}(k)$ are excited states and $f^W(0), 
f^F(0)$ are ground states of the Wannier exciton in dot array and Frenkel 
excitons in medium, respectively. 
Since we will consider only the lowest excited states of the exciton, so 
from now on we will omit the indices $l$ and $l'$. The coefficients u(k) 
and v(k) have the following form:  
\begin{eqnarray} 
u(k) & = &\frac{G(k)}{\{[E^{F}(k) - E^{W}_{tr}(k) ]^2 + G^2(k)\}^{1/2}} \nonumber \\ 
v(k) & = & \frac{E^{F}(k) - E^{W}_{tr}(k) } {\{(E^{F}(k) - E^{W}_{tr}(k)]^2 + G^2(k)\}^{1/2}} 
\end{eqnarray} 

\noindent
Then in term of hybrid
operators $\alpha_{\vec k}, \alpha^+_{\vec k}$, the Hamiltonian in equation (6) can be written: 
\begin{equation} 
H' =\sum_k{E(k)\alpha^+_{\vec k}\alpha_{\vec k}} 
\end{equation}

\noindent
with the energy $E({\vec k})$ of the hybrid state given by the following 
dispersion relation: 
\begin{equation}
E({\vec k}) = 1/2\{E^{F}({\vec k}) + E^{W}_{tr}({\vec k})\} \pm 
1/2\{[E^F({\vec k}) - E^{W}_{tr}({\vec k})]^2 + 
4G^2({\vec k})\}^{1/2} 
\end{equation}

\noindent
Due to the weak dependence of the Frenkel exciton energy upon the k- vector, 
the 
Frenkel exciton energy may be taken independent of the wave vector k,  
$E^F(k) = E^F(0)$.
We can see from (11) that the existence of the array of dots, which 
results in 
the appearance of the transfer exciton energy $E^{W}_{tr}({\vec k})$, enhances 
the coupling between these two kinds of exciton at resonance, i.e., the 
gap between two hybrid exciton branches becomes large. The coupling will be 
strong when $E^F({\vec k})$ and $E^W_{tr}({\vec k})$ are in resonance.
So, the  resonance coupling behavior depends strongly on the hopping 
coefficient  $t({\vec k})$ and the hybridization coefficient $G({\vec 
k})$, which we will investigate in the next sections.

 {\bf III.Hopping coefficient $t({\vec k})$}

The hopping coefficient $t({\vec k})$ is in fact the transfer energy between 
two nearest dots in the dot array and plays an important role in the process 
occurring when a dot array is placed in an organic medium. 
The transfer 
energy is estimated as the electrostatic interaction 
between excitons in dots.
Due to the existence of "confined" 
excitons, each dot has its transition dipole moment, which will interact 
with the corresponding moment of another dot when the distance between 
dots is comparable to the dot radius.
As mentioned above, for one isolated quantum dot the oscillator strength is 
concentrated mainly on the lowest excited states and so we assume that only 
the transition dipole moment to the lowest excited states are involved in 
the interaction for an array.
This multipolar interaction is intrinsically strongly short-range, and
dependent upon the distance between dots,so we use here the nearest 
neighbor approximation.

Now consider the electrostatic interaction between excitons in two spherical 
quantum dots:
\begin{equation}
t({\vec k}) = <W_i({\vec k})| H_{d-d} | W_j({\vec k})>
\end{equation}

\noindent
where $W_i({\vec k}), W_j({\vec k})$ are the exciton wave functions in 
two dots, 
\begin{equation}
|W_i({\vec k})> = 
\frac{1}{V_0}\int{\phi(\vec{r}^{eh}_i)\varphi(\vec{r}_i)
e^{i\vec{k}(\vec{r}_{ei} + \vec{r}_{hi})/2}\Psi^+_e(\vec{r}_{ei})
\Psi^+_h(\vec{r}_{hi})d\vec{r}_{ei}d\vec{r}_{hi}}|0> 
\end{equation}

\noindent
where $\phi(\vec{r}^{eh}_i)$  is the relative electron-hole motion 
function, $\vec{r}^{eh}_i = \vec{r}_{ei} - \vec{r}_{hi}$, 
$\varphi(\vec{r})$ is the exciton envelope function, $\vec{r}_{ei}, 
 (\vec{r}_{hi}), \psi ^+_e(\vec{r}_{ei}), (\psi^+_h(\vec{r}_{hi}))$ are the 
coordinates
and creation operators of electron (hole) in the dot, respectively. 
$V_0$ is the spherical dot's volume.

The envelope function for the exciton inside a spherical dot 
depends on the radius of the dot and has the form [1,4]:
\begin{equation}
\varphi(\vec{r}_i)_{nlm} =  
Y_{lm}(\theta_i,\phi_i)\frac{2^{1/2}}{R^{3/2}}\frac{J_{nl}
(\gamma_{nl}\frac{r_i}{R})}{J_{l+1}(\gamma_{nl})} 
\end{equation}

\noindent 
where $r_{eh}$ is the electron-hole relative 
coordinate and r is the center of mass coordinate of the exciton.
$H_{d-d}$ is the interaction Hamiltonian between two dipole moments 
in these two dots. In our case  the distance between two dots is larger 
than the dot radius $d > R$, so the interaction can be considered 
approximately as the ordinary dipole-dipole interaction. But when the dot 
spacing is of the same order as of the dot radius, we have to consider 
the multipolar interaction. 
Neglecting the higher orders we obtain for the hopping coefficient 
between two spherical quantum dots: 
\begin{equation}
t(k) = \phi_{ns}(0)^2 f_{ns}\{(\vec{\mu}_1^w.\vec{\mu}_2^w) -
3(\vec{\mu}_1^w.\hat{n}_{12})(\vec{\mu}_2^w.\hat{n}_{12})\}
\end{equation}

\noindent $\phi_{nl}(r)$ is the quantum dot exciton envelope function, 
$\vec{\mu}_{1,2}^w$ are transtion dipole moment to the excited state $(n, 
l=0, m=0)$ for the quantum dot spheres 1 and 2, respectively, 
$\hat{n}_{12}$ is the unit vector directed along the straight 
line connecting two excitons, which due to the small dot radius we can 
approximately  treat as directed along the line connecting two 
dot centers. $f_{ns}$ is the integral: 
\begin{equation} 
f_{ns} = \int{\varphi(\vec{r})_{ns}d^3r}\int{\frac{\varphi(\vec{r'})_{ns}} {|d + r + r'|^3}d^3r'} 
\end{equation}

\noindent
The integrals are taken over the volume of the two dots.

For the exciton polarization parallel to the direction connecting two dot 
centers: 
\begin{equation} 
t_{\|} = -\phi_{ns}(0)^2f_{ns}2(\mu^w)^2 
\end{equation}

\noindent
For the exciton polarization perpendicular to the direction connecting
two dot centers the hopping coefficient is equal to:
\begin{equation}
t_{\bot} = \phi_{ns}(0)^2f_{ns}(\mu^w)^2
\end{equation}

\noindent
The hopping constant depends strongly on the polarization
direction of excitons. Also, the transfer energy (7) depends on the 
direction of k-vector and relationship between the k-vector and the 
polarization mode of the exciton. The longitudinal and transverse modes 
have different energies. 

For the transition dipole moment to the excited state $(n, l = 0, m = 0)$
of the spherical quantum dot we have:
\begin{equation}
\mu^w = \frac{(2)^{3/2}}{n\pi}\phi_{1s}(0)p_{cv}R^{3/2}
\end{equation}

\noindent

Hence, from (15) - (19), the hopping coefficient t depends on $R/d$, 
so we can change the dot separation d with respect to dot radius R in 
order to determine the optimum t.

{\bf IV.Hybridization parameter G(k)}

The organic medium can also be described as a lattice, with organic 
molecules occupying every site. The Frenkel exciton can move between the 
sites. Because of the small "lattice constant" here, the organic 
molecular lattice can be considered as a "microscopic" lattice in comparison 
with the macroscopic size of the dot lattice. The organic lattice 
constant is of order $5 \AA $, while the dot radius is about 30-100 
$\AA$ and the dot lattice constant is usually around 60-500 $\AA$.  
The resonance coupling of Frenkel excitons in the medium and Wannier 
excitons in the dot array is determined by the interaction parameter
\begin{equation}
G(k) = <F,k| H_{int}|W,k>
\end{equation}

\noindent
where the interaction Hamiltonian is taken similarly to [1, 3, 4]
\begin{equation}
H_{int} = - \sum_n{E(r_n)P(r_n)}
\end{equation}

\noindent
Here $E(r_n)$ is the operator of the electric field created at point 
$r_n$ in the organic medium by the excitons in quantum dots, $P(r_n)$ is 
the transition 
polarization operator of the Frenkel exciton at molecular site $r_n$ of
the organic medium.

If the dielectric constant of the semiconductor dots is $\epsilon_1$,
and of the organic medium is $\epsilon_2$, the field at some point $r_n$ 
outside of the dot, created by the exciton in the dot located at r is:
\begin{equation}
E(r-r_n) = \frac{3\epsilon_1}{2\epsilon_2 + \epsilon_1}
\frac{3\hat{n}\cos{\theta} - \hat{\mu}}{(r-r_n)^3}\mu^D(\hat{a}_{nl}^+(r) +
\hat{a}_{nl}(r))
\end{equation}

\noindent

$P(r_n) $ is the polarization operator of the Frenkel exciton at molecular 
site $r_n$,
\begin{equation}
P(r_n) = \vec{\mu}^F(\hat{b}^+_n + \hat{b}_n)
\end{equation}

\noindent
$\vec{\mu}^F$ is the optical transition dipole moment of the organic 
molecule. The Frenkel exciton wave function is written in the form:
\begin{equation}
F(k) = \frac{1}{N_F^{1/2}}\sum_n{e^{ikr_n}\chi^f_s(r_n)b^+_n}|0>
\end{equation}

\noindent
$\varphi_s(r_n)$ is the excited state of the molecule at site $r_n$. 
Putting (21) - (24) into (20), we have the expression for
the hybridization coefficient of the semiconductor quantum dot and the
organic medium
\begin{equation}
G(k) = \frac{3\epsilon_1}{2\epsilon_2 + \epsilon _1}\frac{\pi}{2}
\frac{\sin\theta}{(N^F)^{1/2}}\mu^F\mu^w\phi_{ns}(0)D_{ns}(k)
\end{equation}

\noindent
where $\theta$ is the angle between exciton transition dipole moments of
the quantum dot and the organic molecule, and  
\begin{equation}
D_{ns}(k) = 
\int_{Medium}{e^{ikr'}\chi^f_{ns}(r')d^3r'}\int_{Dot}{\frac{\varphi_{nlm}(r)}
{r|r - r'|^3}d^3r}
\end{equation}

\noindent
The first integral is taken over the dot and the second one is taken 
over the volume of the whole medium.

For illustration a numerical calculation was done for some samples.
Fig.1 shows the hybrid exciton dispersion curves plotted for ZnSe 
dots embedded in a standard organic material. The parameters were taken as 
$E^F(0) - E^W(0) = 5 meV, a_B = 30 \AA,\mu^F = 5 D, N = 5.$ 
In Fig 1 two branches of the hybrid exciton are plotted for an array of dots
with radius $ R = 40 \AA$, and the dot lattice constant $d = 80 \AA$. 

{\bf V.Non - Linear Optical Response}.

As already noted in [11] and can be seen from (7), the Wannier transfer 
exciton has a rather small translational mass, which depends on the hopping 
constant and the number of dots. This small translational mass is one 
reason for a large coherence length, which is related to the homogeneous 
linewidth of the excitonic transition. In turn, this 
leads to a large figure of merit of the exciton, 
and, associated with it, large exciton resonance
oscillator strength and nonlinearity [11].

In the region of strong mixing, the oscillator strength of the hybrid 
exciton is determined as:
\begin{equation}
f(\vec{k}) = |u(\vec{k})|^2f^F + |v(\vec{k})|^2f^W + 
2|u(\vec{k})v(\vec{k})|(f^F)^{1/2}(f^W)^{1/2} 
\end{equation}

\noindent
Both components, the transfer Wannier and the Frenkel excitons, 
give contributions to the oscillator strength of the hybrid exciton. Due to 
the confinement effect of the exciton in one single dot as well as the 
transfer 
exciton coupling between dots in the array, the Wannier transfer exciton 
oscillator strength may achieve a value comparable to that of the 
Frenkel exciton, which is very big due to their small exciton radius and 
small molecular lattice constant. So, placing of many semiconductor 
dots in an organic medium leads to a 
large oscillator strength of the hybrid exciton. At resonance, the 
oscillator strength of the hybrid state is determined by its Frenkel 
exciton component.

In the presence of an electric field, the third-order susceptibility 
can be calculated using standard pertubation theory [12-15]. 
We introduce the decay constant $\gamma$ and note that the 
two-body interaction here is considered to arise from the interaction of 
excitons in different dots. 
Considering only 
the resonance case and neglecting contributions from the other nonresonant 
levels we have approximately the result for the lowest optical 
nonlinearity of the hybrid excitons:
\begin{equation}
\chi^{(3)}(w) \approx\frac{{\mu_F}^4}{V}
\frac{(2\sqrt{2})^4}{\pi^2}(\frac{V_{Medium}}{V_{cell}})^{2}
l_c^3(\frac{R}{d})^6\phi_{1s}^4(0)
\times\{\frac{1}{(w-\tilde{w} + i\gamma_{\perp})^2(w-\tilde{w} -
i\gamma_{\|})}\}
\end{equation}

\noindent 
The exciton coherence length $l_c$ is given as [8]:  
\begin{equation} 
l_c = (\frac{3\pi^2}{2^{1/2}})^{1/3}\frac{\hbar}{(M\hbar\Gamma_h)^{1/2}} 
\end{equation} 

\noindent 
M is the exciton translational mass, which is inversely proportional to the 
transfer energy (7), which we can control by changing the system
parameters namely: dot radius and dot spacing. 
$\hbar\tilde{w}$ is the lowest excitation 
energy of the hybrid exciton. $\Gamma_h$ is the linewidth of the exciton 
and $\gamma_{\perp}, \gamma_{\|}$ are the transverse and longitudinal 
relaxation constant of the excitonic transition, respectively. V is 
the volume of the whole system, $V_{Medium}$ is the volume of the 
organic host, and $V_{cell}$ is the volume of one cell in the organic 
lattice. 
Notice here that we consider the case where the sample size is smaller than 
the coherence length, so the size dependence appears as in (29). 

The value of $\chi^{(3)}(w)$ in (28) at resonance may be very large. By 
changing the number of dots and other parameters of the array, one 
can control value of the non-linearity.

Here we present some numerical results for ZnSe dots.

In Fig. 2 the third-order nonlinear susceptibility of ZnSe quantum dot 
lattice embedded in an organic material is 
plotted for several different dot radii and dot spacings. We use here 
the following typical parameters of organic and semiconductor materials: 
$ v^F = 100 \AA, a_{org} = 5 \AA, \mu^F = 5 D, a_{1B} = 30 \AA, E^F(0) - E^W(0) = 
5meV$. We see that 
at resonance, e.g. where the hybridization is strongest, we have very 
high peak of nonlinear susceptibility with the enhancement of about 5 
orders of magnitude in comparison with that of the Wannier exciton. The 
nonlinear 
coefficient is larger for smaller dots and closer spacing between them. 
Also we expect that disorder in the semiconductor dot array will decrease 
the enhancement effect, but this has not yet been calculated.
 
{\bf Summary}.

In summary, we presented here the possibility of creating systems
which offer a strong resonance coupling of Frenkel and Wannier excitons
to obtain an hybrid exciton  state with the special properties of both
kinds of exciton, i.e.having large exciton radius as well as large
oscillator strength. In addition, we can control the expected resonance
parameters by changing the number of dots, their radius, the distance 
between them, and the dot rdius and dot separation relation. Dr. D. Norris 
suggested study of the system with randomness - i.e., the 
disorder distribution of dots, which is very interesting and important. 
We are investigating this.

We are grateful to Professor V. M. Agranovich and Dr. D. Norris for useful 
discussion. We also want to thank a referee for suggestion we use 
translational invariance for the coupling constant. 

\newpage
\begin{center}
{\bf REFERENCES}
\end{center}
\begin{enumerate}
\itemsep -.1cm

\item V. Agranovich, Solid State Comm. {\bf 92}, 295 (1994).  
\item V. I. Yudson, P. Reineker, V. M. Agranovich, Phy. Rev. B, {\bf 52}, R5543 (1995).  
\item A. Engelmann, V. I. Yudson, P. Reineker, Phy. Rev. B, {\bf 57}, 1784 (1998).  
\item V. M. Agranovich, D. M. Basko, G. C. La Rocca, F. Bassani, J. Phys.  
Condens. Matter {\bf 10}, 9369 (1998). 
\item L. Brus , "Semiconductor Colloid: Individual Nanocrystals, Opal
and Porous Silicon" (preprint). \\
   L. E. Brus, J. Chem. Phys. {\bf 80},4403(1982).
\item J. J.Shiang, A.V. Kadavanich, R. K. Grubbs and A. P. Alivisatos,
J. Phys. Chem. {\bf 99}, 17417 (1995). 
\item D. J. Norris, A. Sacra,C. B. Murray,
M. G. Bawendi, Phys. Rev. Lett. {\bf 72}, 2612 (1994), Phys. Rev. B{\bf 
53}, 16338 (1994).
\item A.L. Efros, M. Rosen, M. Kuno, M. Nirmal, D. J. Norris, M. Bawendi, 
   Phys. Rev. B {\bf 54}, 4883 (1996).\\
   M. Nirmal, D. J. Norris, M. Kuno, M. G. Bawendi, A.L. Efros, M. Rosen,
   Phys. Rev. Lett.{\bf 75}, 3728 (1995)\\
   A. L. Efros and A.V. Rodina, Phys.Rev. B {\bf 47}, 10005 (1993). 
   A. L. Efros, Phys. Rev. B{\bf 46}, 7448 (1992).
\item  E. Hanamura, Phys. Rev. B {\bf 37}, 1273 (1988).
\item  C. B. Murray, C. R. Kagan, M. G. Bawendi, Science {\bf 270}, 1335 (1995).
\item T. Takagahara, Solid State Comm.,{\bf 78},No. 4, 279 (1991); Surface 
Science {\bf 267}, 310 (1992).\\
   T. Takagahara and E. Hanamura, Phys. Rev. Lett. {\bf 56}, 2533 (1986).
\item S. Schmitt-Rink, D. A. B. Miller and D. S. Chemla, Phys. Rev. B {\bf 
35}, 8113 (1987).\\
   L. Banyai, Y.Z. Hu, M. Lindberg, S. W. Koch, Phy. Rev. B {\bf 38}, 
8142 (1988).
\item Karl W. Boer, "Survey of Semiconductor Physics: 
Electron and other Particles in Bulk Semiconductors", Van Nostran 
Reinhold, New York 1990. 
\item T. S. Moss, Handbook on semiconductors, vol.1,p. 23, 
North-Holland Publishing Company, Amsterdam-NewYork-Oxford 1982.
\item S.Mukamel, Principle of Nonlinear Optical Spectroscopy,
Oxford University Press, New York-Oxford 1995.
\item S. Facsko, T. Dekorsy, C. Koerdt, C. Trape, H. Kurz, A. Vogt, H. L. 
Hartnagel, Science {\bf 285}, 1551 (1999). 
\end{enumerate}

\newpage
{\bf FIGURE CAPTIONS}
\vspace{2cm}

\noindent
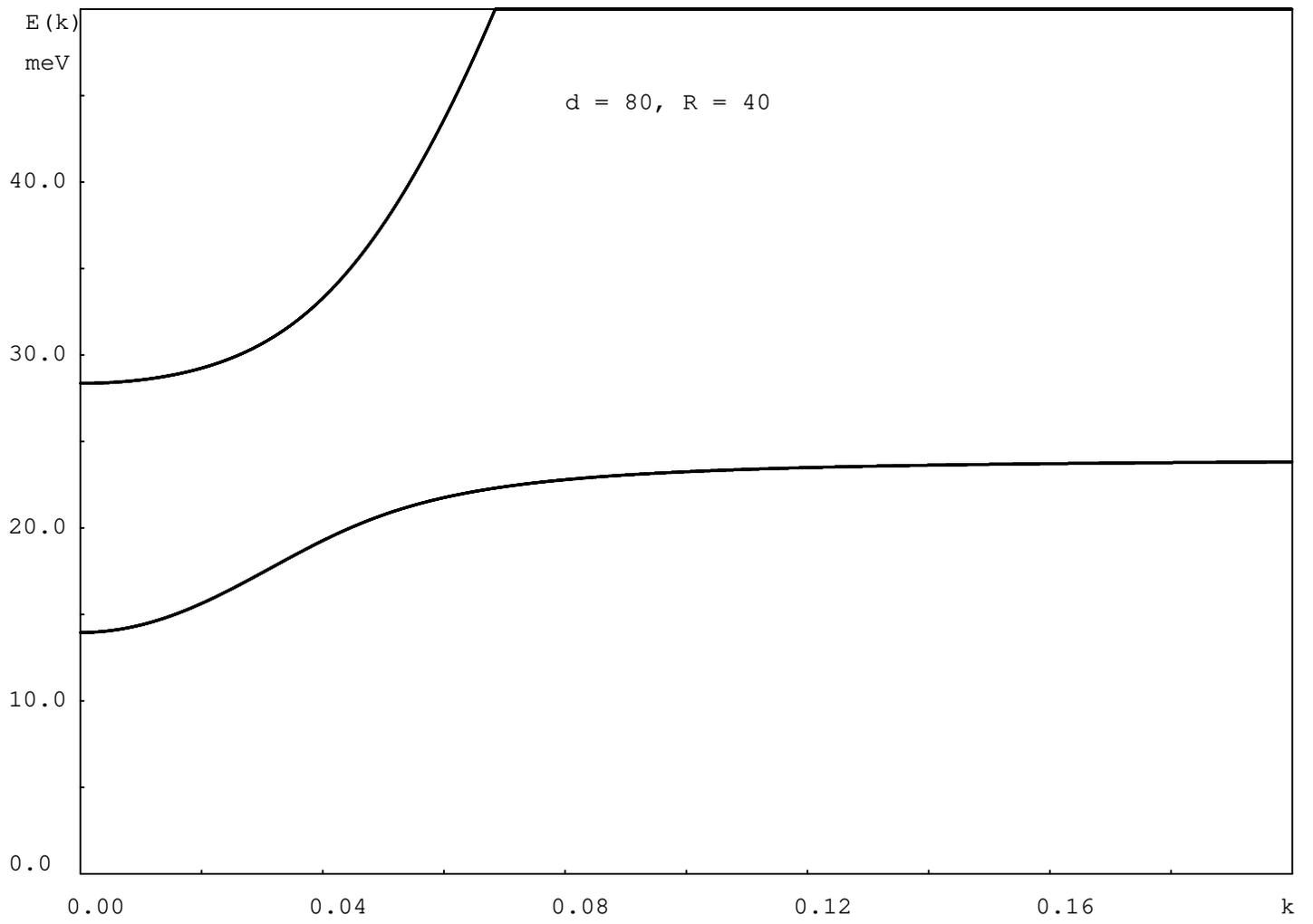
{\bf Fig.1}.Hybrid Exciton Dispersion calculated for ZnSe dots placed in 
an organic material. The dispersion curve is ploted for dot radius $R=50 A$
and dot spacing $d=200 A$. \\ 
\vspace{0.7cm} 

\noindent
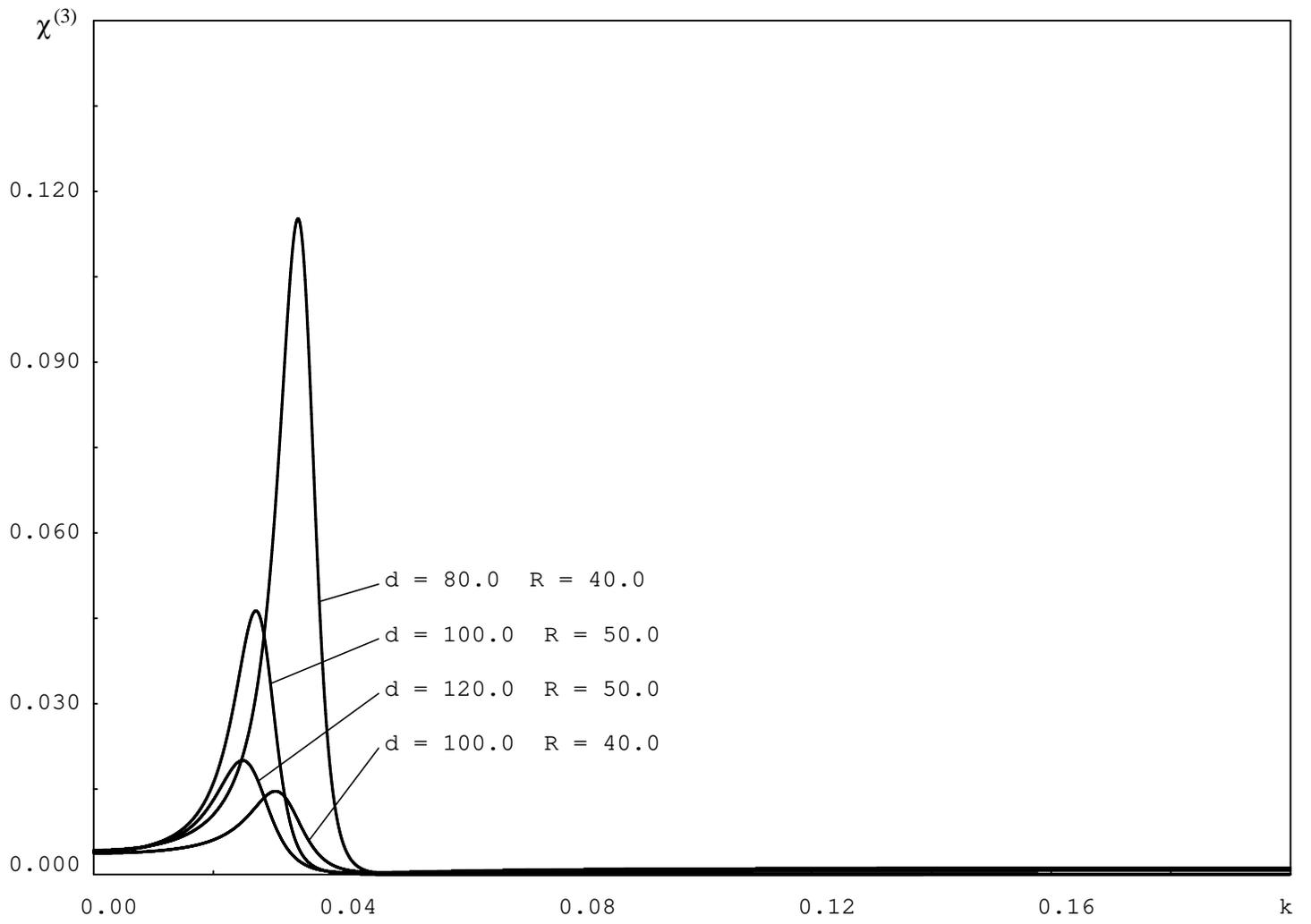
{\bf Fig.2}.Third-order nonlinear susceptibility for the hybrid exciton 
state of ZnSe dots in organic material.WE plot the susceptibility versus 
wave - vector. Note that there is a very significant enhancement of 
$\chi^{(3)}$. Plots are for different dot radius $R$ and separation $d$. \\ 

\end{document}